\documentclass[12pt]{article}
\usepackage{epsfig}
\def\u1p{U(1)$^{\prime}$}
\def\zp{$Z^{\prime}$ }
\newcommand{\be}{\begin{equation}}
\newcommand{\ee}{\end{equation}}
\newcommand{\bea}{\begin{eqnarray}}
\newcommand{\eea}{\end{eqnarray}}
\newcommand{\wtd}{\widetilde}
\newcommand{\bth}{{\bf 3}}
\newcommand{\btw}{{\bf 2}}
\newcommand{\bon}{{\bf 1}}
\def\simlt{\mathrel{\raise.3ex\hbox{$<$\kern-.75em\lower1ex\hbox{$\sim$}}}}
\def\simgt{\mathrel{\raise.3ex\hbox{$>$\kern-.75em\lower1ex\hbox{$\sim$}}}}
\begin{document}

\begin{titlepage}
\begin{flushright}
MCTP-05-69\\
IZTECH-P05/01
\end{flushright}
\vspace{0.5cm}
\begin{center}
{\Large \bf The minimal \u1p extension of the MSSM} \\
\vspace{1cm} \renewcommand{\thefootnote}{\fnsymbol{footnote}}{\large 
Durmu{\c s} A. Demir$^1$\footnote{Email: demir@physics.iztech.edu.tr}, 
Gordon L. Kane$^2$\footnote{Email: gkane@umich.edu} and 
Ting T. Wang$^2$\footnote{Email: tingwang@umich.edu}
\renewcommand{\thefootnote}{\arabic{footnote}} \\ {\it
1. Department of Physics, Izmir Institute of Technology, \\
Izmir,  TR 35430, Turkey \\
2. Michigan Center for Theoretical Physics \\ Ann
  Arbor, MI 48109, USA}}
\end{center}
\vspace{0.5cm}


\begin{abstract}
Motivated by the apparent need for extending the MSSM and perhaps mitigating 
naturalness problems associated with the $\mu$ parameter and fine-tuning
of the soft masses, we augment the MSSM spectrum by a SM gauge singlet 
chiral superfield, and enlarge the gauge structure by an additional \u1p
invariance, so that the gauge and Higgs sectors are relatively
secluded. One crucial aspect of \u1p models is the existence of
anomalies, cancellation of which may require the inclusion of exotic
matter which in turn disrupts the unification of the gauge couplings. In
this work we pursue the question of canceling the anomalies with a
minimal matter spectrum and no exotics. This can indeed be realized
provided that \u1p charges are family-dependent and the soft-breaking
sector includes non-holomorphic operators for generating the fermion
masses. We provide the most general solutions for  \u1p charges by
taking into account all constraints from gauge invariance and anomaly
cancellation. We analyze various laboratory and astrophysical bounds
ranging from fermion masses to relic density, for an illustrative set of
parameters. The \u1p charges admit patterns of values for which family
nonuniversality resides solely in the lepton sector, though this does
not generate leptonic FCNCs due to the \u1p gauge invariance.
\end{abstract}
\vspace*{1cm}
\end{titlepage}

\section{Introduction}
Supersymmetric models extending the minimal supersymmetric model
(MSSM) are generally motivated for stabilizing the $\mu$ parameter
at the electroweak scale, and for incorporating right-handed
neutrinos into the spectrum. The extension of the MSSM may or may
not involve additional gauge groups. Concerning the former, the
most conservative approach is to extend the gauge structure of the
MSSM by an extra Abelian group factor \u1p along with an
additional chiral superfield $\widehat{S}$ whose scalar component
generates an effective $\mu$ parameter upon spontaneous \u1p
breakdown. The \u1p symmetry in question is essentially the
gauging of the global Peccei-Quinn invariance of the MSSM. What it
actually does is to forbid a bare $\mu$ parameter thereby
providing a dynamical solution to the $\mu$ problem \cite{muprob}.
Extra U(1) symmetries arise as low-energy manifestations of
grand unified \cite{gut}, of string \cite{string}, and of
dynamical electroweak breaking \cite{dynamic} theories.

An important property of \u1p models is that the lightest Higgs
boson weighs significantly more than $M_Z$ even at tree level with small
$\tan\beta$. Hence
the existing LEP bounds are satisfied with almost no need for large
radiative corrections \cite{Cvetic:1997ky,everett,han}. Besides,
they offer a rather wide parameter space for facilitating the
electroweak baryogenesis \cite{tianjun}.

An important issue about extra \u1p models concerns the
cancellation of anomalies. Indeed, for making the theory
anomaly--free the usual approach to \u1p models is to
add several exotics to the spectrum \cite{Erler:2000wu}.
This not only causes a significant departure
from the minimal structure but also disrupts the gauge coupling
unification -- one of the fundamental predictions of the MSSM with weak
scale soft masses.

The prime goal of the present work is to construct an anomaly-free
\u1p model without exotics. We accomplish this by
allowing family-nonuniversal \u1p invariance. It is known that
when different fermion families posses different \u1p charges
generally large \zp--mediated flavor-changing neutral currents
(FCNC) arise \cite{Langacker:2000ju}. However, there are
exceptions to this, especially when \zp FCNC effects reside in the
lepton sector. For example, if the \u1p charges forbid the off-diagonal
terms in the fermion mass matrix (in the family space), the mass
eigenstates will coincide with the gauge eigenstates. Therefore, there
will be no FCNC induced by the \zp gauge boson.
The family-dependence of the \u1p invariance necessarily forbids
certain Yukawa couplings in the superpotential, leading to
massless fermions. The requisite fermion masses, however, can be
induced at the loop level via non-holomorphic operators in the soft sector
\cite{Hall:1990ac,Borzumati:1999sp}. In addition to being allowed, these
non-holomorphic terms can appear in
intersecting brane models with certain types of fluxes turned on
\cite{Camara:2003ku}. Therefore, as we will describe in the text, a
minimal \u1p model can be realized with family-dependent charges
and non-holomorphic terms.

The paper is organized as follows. In section 2 below we introduce
non-holomorphic terms and discuss  how the fermion masses as well as other
chirality-changing operators such as the magnetic moments are
induced. In section 3 we discuss in detail the construction of an
anomaly-free \u1p model with minimal matter content. We also
determine the flavor structures of the Yukawa matrices and of the
non-holomorphic terms therein. In section 4 we survey phenomenological tests of the
\u1p models by briefly discussing fine tuning, the Higgs sector, \zp couplings,
collider signatures, neutrino masses, muon $g-2$, and the relic
density of the universe. In section 5 we conclude the work.

\section{\u1p Models with non-holomorphic SUSY breaking}
\label{zp:sec:cte}
In \u1p models the MSSM gauge group is extended to include an extra Abelian
group factor: SU(3)$_c\times$SU(2)$_L \times$U(1)$_Y\times$U(1)$^{\prime}$ with
respective gauge couplings $g_3$, $g_2$, $g_Y$ and $g_1^{\prime}$.
This gauge structure survives all the energy scales from $M_{GUT}\approx
2\times 10^{16}\, {\rm GeV}$ down to a ${\rm TeV}$. The particle spectrum of
the model is that of the MSSM plus a MSSM gauge singlet $S$ charged
under only the \u1p invariance. Clearly, the family-universality of the
MSSM gauge charges
is not necessarily respected by the \u1p group. Hence we employ a
general family-dependent charge assignment as tabulated in
Table \ref{zp:tab:qn}.
\begin{table}[h]
\begin{center}
\begin{tabular}{|c|c|c|c|c|} \hline
      & SU(3)$_c$ & SU(2)$_L$  & U(1)$_Y$ & U(1)$^{\prime}$   \\ \hline
$Q_i$   & $\bth$       &  $\btw$       &  $1/6$   & $Q_{Q_i}$ \\ \hline
$U^c_i$   & $\bar \bth$  &  $\bon$       &  $-2/3$  & $Q_{U^c_i}$ \\ \hline
$D^c_i$   & $\bar \bth$  &  $\bon$       &  $1/3$   & $Q_{D^c_i}$ \\ \hline
$L_i$   & $\bon$       &  $\btw$       &  $-1/2$  & $Q_{L_i}$ \\ \hline
$E^c_i$   & $\bon$       &  $\bon$       &  $1$     & $Q_{E^c_i}$ \\ \hline
$H_u$ & $\bon$       &  $\btw$       &  $1/2$   & $Q_{H_u}$ \\ \hline
$H_d$ & $\bon$       &  $\btw$       &  $-1/2$  & $Q_{H_d}$ \\ \hline
$S$   & $\bon$       &  $\bon$       &  $0  $   & $Q_{S}$ \\ \hline
\end{tabular}
\end{center}
\caption{The gauge quantum numbers of chiral fields in the \u1p model. The
index $i$ runs over three families of matter. Each family can
acquire a different charge under the \u1p group.}
\label{zp:tab:qn}
\end{table}
\newline
The superpotential takes the form:
\be \widehat{W}= h_s \widehat{S} \widehat{H}_d
\widehat{H}_u + h_u^{ij} \widehat{U}^c_{j} \widehat{Q}_i
\widehat{H}_u + h_d^{ij} \widehat{D}^c_j \widehat{Q}_i
\widehat{H}_d+ h_e^{ij} \widehat{E}^c_j \widehat{L}_i
\widehat{H}_d \label{zp:eq:supo}
\ee
The first term of the superpotential induces an
effective $\mu$ parameter $h_s \langle S \rangle$ below the scale
of \u1p breaking. This provides a dynamical solution to the $\mu$
problem when $\langle S \rangle \sim {\cal{O}}({\rm TeV})$. The
rest of the operators in (\ref{zp:eq:supo}) describes the Yukawa
interactions of leptons and quarks.

The most general holomorphic structures which break supersymmetry softly are
\bea \label{soft}
\nonumber
-{\mathcal{L}}_{soft}&=&\left(\sum_i M_i \lambda_i\lambda_i- A_Sh_sSH_dH_u
 -A_u^{ij}h_u^{ij} U^c_jQ_iH_u-A_{d}^{ij}h_d^{ij} D^c_jQ_iH_d-
 A_e^{ij}h_e^{ij} E^c_jL_iH_d+h.c.\right)  \\ \nonumber
 &+& m_{H_u}^2|H_u|^2+m_{H_d}^2|H_d|^2+m_S^2|S|^2+  \\
 &+&  m_{Q_{ij}}^2 \wtd Q_i\wtd Q_j^*
    +m_{U_{ij}}^2 \wtd U^c_i\wtd U^{c*}_j+m_{D_{ij}}^2 \wtd D^c_i\wtd D^{c*}_j
    +m_{L_{ij}}^2 \wtd L_i\wtd L_j^*+m_{E_{ij}}^2 \wtd E^c_i\wtd E^{c*}_j+h.c.
\eea
where the sfermion mass-squared $m_{Q,\dots,E^c}^2$ and the trilinear
couplings $A_{u,\dots,e}$ are $3\times 3$ matrices in flavor space. All
these soft masses will be taken here to be diagonal.
Moreover, all gaugino masses $M_i$ and trilinear
couplings $A_{S, \dots, e}$ will be taken real since the
(important and interesting) question of CP violation is
beyond the scope of the present work (interested readers can refer to \cite{everett}).

Clearly, the \u1p charge assignments of chiral superfields put
stringent constraints on the Yukawa textures \cite{Jack:2003pb}. 
For instance, if the
\u1p charges satisfy \be Q_{Q_1}+Q_{U_i}+Q_{H_u}\ne 0
\qquad\mbox{for}\qquad i=1,2,3 \ee then the up quark can acquire a
mass neither at tree level nor at any loop level with holomorphic
soft terms. Therefore, for avoiding massless fermions it is
necessary to introduce non-holomorphic SUSY-breaking operators,
the non-holomorphic terms \cite{Hall:1990ac,Borzumati:1999sp,
Camara:2003ku,ma1,jack}. Historically, the non-holomorphic terms
have not been classified as 'soft' since they might give rise to
quadratic divergences \cite{nonholo}. However, such operators are
perfectly soft when no gauge singlets are contained in the theory.
Indeed, non-holomorphic terms are soft in the MSSM and its \u1p
extensions. Concerning the origin of the non-holomorphic terms,
one notes that they are generated by spontaneous SUSY breaking
within gravity mediation \cite{martinex}. In addition to this,
they arise naturally in  strongly coupled SUSY gauge theories
\cite{nima}. Moreover, the effective potentials of $N=2$ and $N=4$
SUSY gauge theories are endowed with radiatively-generated
non-holomorphic soft terms \cite{buch}.

For the \u1p model under concern the non-holomorphic SUSY breaking lagrangian
takes the form \be \label{mssmnonholo} -{\mathcal{L}}_{c}=
C_E^{ij}H_{u}^* \wtd L^i \wtd E_R^{cj}+
 C_U^{ij} H_{d}^* \wtd Q^i\wtd U_R^{cj} +
 C_D^{ij} H_{u}^* \wtd Q^i\wtd D_R^{cj} + c.c.
\ee and need to be added to the holomorphic ones in (\ref{soft}).
Clearly, a down-type quark, for instance, develops a finite mass
via triangular diagrams proceeding with  $\wtd D_L$ $\wtd D_R^c$
and a neutral gaugino $\lambda$, and the result is necessarily
proportional to $C_D$. This radiative induction of the fermion
masses is rather generic. Notice that coupling to the 'wrong'
Higgs doublet in (\ref{mssmnonholo}) is essential for giving mass
to fermions. Indeed, a fermion $f$ obtains the mass
\cite{Borzumati:1999sp,murayama} \be \label{fmass} m_f=(C_f\,
v_{\alpha})\left[\frac{\alpha_s}{2\pi} \xi_f m_{\wtd g}
 I_m(m^2_{\wtd f_1},m^2_{\wtd f_2},m^2_{\wtd g})+
 \frac{\alpha_Y}{2\pi}\sum_{j=1}^6 K^j_f m_{\wtd \chi^0_j}
 I_m\left(m^2_{\wtd f_1},m^2_{\wtd f_2},m^2_{\wtd \chi^0_j}\right)\right]
\ee
where $v_\alpha=\langle H_u\rangle (\langle H_d\rangle)$ for down-type
(up-type) fermions. Here the first term refers to SUSY-QCD contribution
($\xi_f= 4/3\,, 0$ for quarks and  leptons, respectively), and the second term
summarizes the contributions of all neutral Higgsinos and gauginos.
$C_f$ is the corresponding non-holomorphic terms in (\ref{mssmnonholo}),
and $\alpha_Y=g^2_Y/(4\pi)$.
The triangular loop function $I_m$ is defined by
\be
I_m\left(m_1^2,m_2^2,m_\lambda^2\right)= \frac{1}{m_1^2 - m_2^2} \, \left(\frac{\ln\beta_1}{\beta_1-1} - \frac{\ln\beta_2}{\beta_2-1}\right)
\ee
where $\beta_i = m_{\lambda}^2/m_i^2$ with $i=1,2$. This function
approaches $1/2 m^2$
when $m_1 \sim m_2 \sim m_{\lambda}\equiv m$. The coupling of $j$--th
neutralino to mass-eigenstate sfermions (${\wtd f_i}$ with masses
$m_{\wtd f_i}$) is given by
\be
\label{kfj}
K^j_f=\left[Y_{f_R}N_{jB}+\left(\frac{g_1'}{g_Y}\right)
 Q_{f_R}N_{jB'}\right]
 \left[Y_{f_L}N_{jB}+\left(\frac{g_1'}{g_Y}\right)Q_{f_L}N_{jB'}
   +\cot\theta_W N_{jW} T_{3f_L}\right]
\ee where $Q_f$ is the \u1p charge of the fermion $f$, $Y_f =
Q_{em}^f - T_3^{f}$, and $g_1'$ and $g_Y$ stand for the \u1p and
hypercharge gauge couplings, respectively. Here $N_{jB'}$,
$N_{jB}$ and $N_{jW}$ are the \zp, bino and wino components of the
$j$--th neutralino. Note that the fermion masses in (\ref{fmass})
are of the form $m_f = \kappa_f\, \langle H_{\alpha} \rangle$
where the dimensionless coupling in front involves gauge couplings
and sparticle mixing angles as well as the ratios of the trilinear
couplings to sparticle masses. Hence, various soft-breaking
parameters must conspire to generate fermion masses in agreement
with experiment. It might be useful to dwell on this point
briefly. For reproducing the correct hierarchy of the light
fermion masses ($i.e.$ $m_u<m_d$, $m_s<m_c$, $m_e<m_{\mu}$) one
can tune the sfermion masses, the non-holomorphic trilinear
couplings $C_f$ or the \u1p charges. As a simple case study let us
examine the $u$--$d$ mass hierarchy in the limit of degenerate
$\wtd u$ and $\wtd d$ squarks. One finds
\begin{eqnarray}
\frac{m_u}{m_d} = \frac{C_U^{11}}{C_D^{11}}\; \frac{ 1+ \frac{3 \alpha_Y}{4 \alpha_s} \sum_{j =1}^{6}
K_{u}^{j} R_j}{ 1+ \frac{3 \alpha_Y}{4 \alpha_s} \sum_{j =1}^{6}
K_{d}^{j} R_j}
\end{eqnarray}
where $R_j \equiv m_{\wtd \chi^0_j}
 I_m\left(m^2_{\wtd f_1},m^2_{\wtd f_2},m^2_{\wtd \chi^0_j}\right) /
m_{\wtd g} I_m(m^2_{\wtd f_1},m^2_{\wtd f_2},m^2_{\wtd g})\sim
m_{\wtd \chi^0_j}/m_{\wtd g}$ is identical for up and down
squarks. In case $C_U^{11}\simeq C_D^{11}$ the $u$--$d$ hierarchy
can be saturated if $\sum_j (0.5 K_d^j - K_u^j) R_j \sim 10$ which
is too large to be satisfied unless gluino is exceedingly light,
$m_{\wtd g}\sim 1\ {\rm GeV}$. Other fermion masses can be
analyzed in a similar way. Therefore, the hierarchy among the
fermion masses rests largely on the hierarchy of the
non-holomorphic trilinears. On the other hand, generation of the
correct values of the individual fermion masses requires a
judicious choice of the soft masses and \u1p couplings.

As was shown in  \cite{Borzumati:1999sp}, it is difficult to generate
masses for the top quark and tau lepton if the non-holomorphic terms are 
not much larger than
the other soft masses. Therefore, the \u1p charge assignments must be such
that these fermions can obtain masses already at tree level. However,
the rest of the fermions can acquire masses through (\ref{fmass}) with no
obvious contradiction with experiments.

The sparticle virtual effects which give rise to nonvanishing
fermion masses (\ref{fmass}) induce also chirality-violating
operators pertaining to radiative transitions of the fermions.
Among these are the electric and magnetic dipole moments. In fact,
for a fermion with radiatively induced mass the magnetic dipole
moment takes the form \cite{Borzumati:1999sp} \be \label{magmom}
a^{\mathit{SUSY}}_f=2m_f^2\frac{\sum_j K^j_f
m_{\wtd\chi^0_j}I_{g-2}
   \left(m_{\wtd f_1}^2,m_{\wtd f_2}^2,m_{\wtd\chi_j^0}^2\right)}
{\sum_j K^j_f m_{\wtd\chi^0_j}I_m
   \left(m_{\wtd f_1}^2,m_{\wtd f_2}^2,m_{\wtd\chi_j^0}^2\right)} \label{zp:eq:gmu}
\ee
where
\be
I_{g-2}\left(m_1^2,m_2^2,m_\lambda^2\right)=
\frac{1}{m_\lambda^2}\frac{1}{m_2^2-m_1^2}
\left\{\frac{\beta_1(\beta_1^2-1-2\beta_1\log\beta_1)}
           {2(\beta_1-1)^2}-(1\to 2)\right\}
\ee with the same parameterization used for $I_m$. If $\wtd
m=\mbox{max}(m_{\wtd f_1},m_{\wtd f_2},m_{\wtd \lambda})$ then
\be a^{\mathit{SUSY}}_f = 2 m_f^2 \frac{\sum K^j_f
m_{\wtd\chi^0_j}I_{g-2}
   \left(m_{\wtd f_1}^2,m_{\wtd f_2}^2,m_{\wtd\chi_j^0}^2\right)}
{\sum K^j_f m_{\wtd\chi^0_j}I_m
   \left(m_{\wtd f_1}^2,m_{\wtd f_2}^2,m_{\wtd\chi_j^0}^2\right)}
\sim \frac{m_f^2}{3\wtd m^2} \ee so that larger the heaviest
sparticle mass smaller the magnetic moment. One notes that the
expression of the magnetic moment (\ref{magmom}) contains no loop
suppression factor $1/(4 \pi)^2$ due to the fact that fermion mass
itself is generated radiatively. Hence, when the fermion mass is
generated solely by non-holomorphic soft terms the magnetic
moment, in particular the muon magnetic moment $a_{\mu}$, tends to
be large. The most stringent bound is from the measured $a_\mu$. Indeed,
if the muon mass follows from non-holomorphic terms (as will be the case
in our model mentioned below) then for saturating the existing
experimental bounds on $g_{\mu} -2$ the scalar muon $\wtd\mu$ must
weigh $\mathcal{O}(\mbox{TeV})$.

\section{An Anomaly-free Minimal \u1p Model}\label{zp:sec:lag}
One of the most important issues in \u1p models is the
cancellation of gauge and gravitational anomalies. Indeed, for
making the theory anomaly-free one has been forced to augment the
minimal spectrum by a number of exotics \cite{Erler:2000wu}.
These additional fields usually disrupt the unification of
the gauge couplings. In this
section we will discuss the crucial role played by
family-dependent \u1p charges in cancelling the anomalies and
hence in preserving the unification of gauge forces.

For the theory to be anomaly-free the \u1p charges of chiral fields must satisfy
\bea \label{eqn:ano1}
0&=&\sum_i(2Q_{Q_i}+Q_{U^c_i}+Q_{D_i}) \\
0&=&\sum_i(3Q_{Q_i}+Q_{L_i})+Q_{H_d}+Q_{H_u} \\
0&=&\sum_i(\frac{1}{6}Q_{Q_i}+\frac{1}{3}Q_{D^c_i}+
 \frac{4}{3}Q_{U^c_i}+\frac{1}{2}Q_{L_i}+Q_{E^c_i})+
 \frac{1}{2}(Q_{H_d}+Q_{H_u}) \\
0&=&\sum_i(6Q_{Q_i}+3Q_{U^c_i}+3Q_{D^c_i}+2Q_{L_i}+Q_{E^c_i})+2Q_{H_D}
 +2Q_{H_u}+Q_s \\ \label{eqn:ano5}
0&=&\sum_i(Q_{Q_i}^2+Q_{D^c_i}^2-2Q_{U^c_i}^2-Q_{L_i}^2+Q_{E^c_i}^2)-
 Q_{H_d}^2+Q_{H_u}^2 \\ \label{eqn:ano6}
0&=&\sum_i(6Q_{Q_i}^3+3Q_{D^c_i}^3+3Q_{U^c_i}^3+2Q_{L_i}^3+Q_{E_i}^3)+
 2Q_{H_d}^3+2Q_{H_u}^3+Q_S^3
\eea
which correspond to vanishing of $U(1)^{\prime}$-$SU(3)$-$SU(3)$, $U(1)^{\prime}$-$SU(2)$-$SU(2)$,
$U(1)^{\prime}$-$U(1)_Y$-$U(1)_Y$, $U(1)^{\prime}$-graviton-graviton,
$U(1)^{\prime}$-$U(1)^{\prime}$-$U(1)_Y$ and \u1p-\u1p-\u1p anomalies, respectively.

As mentioned before, the top quark and tau lepton masses must be
generated already at tree level. Moreover, \u1p invariance must
allow for $S H_d H_u$ coupling for solving the $\mu$ problem.
These conditions lead to: \bea \label{extracond}
Q_{Q_3}+Q_{U^c_3}+Q_{H_u}&=&0 \\
Q_{L_3}+Q_{E^c_3}+Q_{H_d}&=&0 \\
Q_{H_u}+Q_{H_d}+Q_{S}&=&0 \eea which should be added to eq.
(\ref{eqn:ano1}-\ref{eqn:ano6}). The family-nonuniversal \u1p
charges could lead to large \zp--mediated FCNCs
\cite{Langacker:2000ju}. One first observes
that the very presence of the CKM matrix implies that the physical
quark states are achieved after a unitary rotation of the
gauge-basis quarks. Therefore, for guaranteeing the suppression of
FCNCs in the hadron sector it is good to keep quark \u1p charges
family-universal: \be \label{quarkcharge} \label{eqn:fam}
Q_{Q_1}=Q_{Q_2}=Q_{Q_3}\;, \quad Q_{U^c_1}=Q_{U^c_2}=Q_{U^c_3}\;,
\quad Q_{D^c_1}=Q_{D^c_2}=Q_{D^c_3} \ee so that, depending on the
charge assignments of the Higgs doublets, either down or up quark
sector possesses tree level Yukawa interactions. For the lepton
sector one can relax the condition of family-universality since it
will lead to FCNCs only if mass-- and gauge--eigenstate leptons
are not identical. As will be seen below, \u1p charges can be
assigned in such a way that the mass matrix of leptons is
automatically flavor-diagonal and hence leptonic FCNCs are absent.

We now want to illustrate the assignment of \u1p charges. There
are 18 unknowns and 15 constraints (\ref{eqn:ano1}-\ref{eqn:fam})
out of which (\ref{eqn:ano5},\ref{eqn:ano6}) are nonlinear in
charges. Using the linear constraints we first express 13 charges
in terms of 5 charges which we choose to be \be Q_{L_2}\;,\quad
Q_{E^c_2}\;,\quad Q_{E^c_3}\;, \quad Q_{H_d}\quad\mbox{and}\quad
Q_S\,.\ee The explicit expressions for charges read as \bea
\label{eqn:QQ1}
&&Q_{Q_1}=Q_{Q_2}=Q_{Q_3}=\frac{1}{9}(-3Q_{H_d}-2Q_S) \\
&&Q_{D^c_1}=Q_{D^c_2}=Q_{D^c_3}=\frac{1}{9}(-6Q_{H_d}-7Q_S) \\
&&Q_{U^c_1}=Q_{U^c_2}=Q_{U^c_3}=\frac{1}{9}(12Q_{H_d}+11Q_S) \\
&&Q_{L_1}=Q_{E^c_3}-Q_{L_2}+4Q_{H_d}+3Q_S \\
&&Q_{L_3}=-Q_{E^c_3}-Q_{H_d} \\
&&Q_{E^c_1}=-Q_{E^c_2}-Q_{E^c_3}-6Q_{H_d}-5Q_S \\ \label{eqn:QHU}
&&Q_{H_u}=-Q_{H_d}-Q_S \eea from which it follows that, for all
$i,j$, $Q_{Q_i}+Q_{D^c_j}+Q_{H_d}=-1$ and
$Q_{Q_i}+Q_{D^c_j}-Q_{H_u}=0$. Hence, all of the down quarks get
their masses from non-holomorphic terms via (\ref{fmass}); they
are not allowed to possess Yukawa structures $h_d^{i
j}$ in the superpotential. On the other hand, the up quarks obtain
their masses from superpotential couplings only. Consider now the
muon mass term. There are two alternatives\footnote{When the
determinant of a matrix is non-zero it can not have a row or column
with all zeroes. In fact, one can employ a rotation in the space
of families to make all diagonal entries of the matrix nonzero.
Hence, in the following we will assume that such a rotation has
already been done such that whenever the determinant of a matrix
is nonzero then no diagonal entry can vanish. In particular, one
can employ a family redefinition to make (2,2) element of $h_l$
nonzero.}: \bea && \mbox{either}\quad Q_{L_2}+Q_{E^c_2}+Q_{H_d}=0
 \quad \mbox{ or } \quad
Q_{L_2}+Q_{E^c_2}-Q_{H_u}=0\;.\label{eqn:muon}  \eea The first
option implies that the muon mass follows
entirely from the Yukawa couplings. On the other hand,
the second option restricts muon mass to follow from
non-holomorphic terms only. From (\ref{eqn:QQ1}-\ref{eqn:QHU}),
one can check that the first option leads to
$Q_{L_1}+Q_{E^c_1}+Q_{H_d}=-2Q_S$ and
$Q_{L_1}+Q_{E^c_1}-Q_{H_u}=-Q_S$. However, for solving the $\mu$
problem $Q_S$ must be nonzero, and this implies that electron is
forbidden to acquire its mass from both the Yukawa couplings and
non-holomorphic terms. Hence, this option must be discarded; the muon cannot
develop a mass
from Yukawa couplings. The remaining alternative
implies that $Q_{L_1}+Q_{E^c_1}-Q_{H_u}=0$ so
that both muon and electron receive their masses from
non-holomorphic terms via (\ref{fmass}). Using (\ref{eqn:QHU}) and
the second option in (\ref{eqn:muon}) it is easy to solve for $Q_{H_d}$: \be
Q_{H_d}=-Q_s-Q_{E^c_2}-Q_{L_2}, \ee so that 14 out of 18 charges
get expressed in terms of \be Q_{L_2}\;,\quad Q_{E^c_2}\;,\quad
Q_{E^c_3}\quad \mbox{and}\quad Q_S. \ee With the solutions
obtained so far, the two nonlinear anomaly cancellation
conditions, (\ref{eqn:ano5}) and (\ref{eqn:ano6}), reduce to \bea
&&0=-2(2Q_{E^c_2}-Q_{E^c_3}+2Q_{L_2}+Q_S)\times Q_S  \\
&&0=-3(2Q_{E^c_2}-Q_{E^c_3}+2Q_{L_2}+Q_S)\times \\ \nonumber
&& ( 3 Q_{E^c_2}^2-Q_{E^c_2}Q_{E^c_3}+10Q_{E^c_2}Q_{L_2}
 -2  Q_{E^c_3}Q_{L_2} +8 Q_{L_2}^2+3Q_{E^c_2}Q_S+Q_{E^c_3}Q_S
 +4Q_{L_2}Q_S)
\eea which are simultaneously satisfied when \be
2Q_{E^c_2}-Q_{E^c_3}+2Q_{L_2}+Q_S=0\ee holds. One can eliminate
$Q_{E^c_3}$ from this relation. Then 15 out of 18 charges get
expressed in terms of the three independent ones, \be
Q_{L_2}\;,\quad Q_{E^c_2}\quad \mbox{ and }\quad Q_S,
\label{eqn:free} \ee via the relations \bea
&&Q_{Q_1}=Q_{Q_2}=Q_{Q_3}=\frac{1}{9}(3Q_{E^c_2}+3Q_{L_2}+Q_S) \\ \nonumber
&&Q_{D^c_1}=Q_{D^c_2}=Q_{D^c_3}=\frac{1}{9}(6Q_{E^c_2}+6Q_{L_2}-Q_S) \\ \nonumber
&&Q_{U^c_1}=Q_{U^c_2}=Q_{U^c_3}=\frac{1}{9}(-12Q_{E^c_2}-12Q_{L_2}-Q_S)
\\ \nonumber
&&Q_{L_1}=-2Q_{E^c_2}-3Q_{L_2}\\ \nonumber
&&Q_{L_3}=-Q_{E^c_2}-Q_{L_2} \\ \nonumber
&&Q_{E^c_1}=3Q_{E^c_2}+4Q_{L_2} \\ \nonumber
&&Q_{E^c_3}=2Q_{E^c_2}+2Q_{L_2}+Q_S \\ \nonumber
&&Q_{H_d}=-Q_{E^c_2}-Q_{L_2}-Q_S \\ \nonumber
&&Q_{H_u}=Q_{E^c_2}+Q_{L_2}  \label{noanomal}
\eea with which the theory becomes
completely anomaly free. One can analyze all physical quantities
of interest in terms of three free charges $Q_{L_2}$, $Q_{E^c_2}$
and $Q_S$ without disrupting the unification of gauge
couplings.

The \u1p charges entirely determine the Yukawa textures: they decide which flavors
receive their masses at tree level and which ones at the loop level.
In fact, the flavor structures of the Yukawa matrices can be
determined via the charge matrices of the associated operators:
\begin{eqnarray}
\label{yukawatext}
 \left( Q_{Q_i}+Q_{U^c_j}+Q_{H_u}\right)&=&
 \left(\begin{array}{ccc}
  0 & 0& 0 \\
  0 & 0& 0 \\
  0 & 0& 0 \\
  \end{array} \right)_{ij}\;,\nonumber\\
 \left( Q_{Q_i}+Q_{D^c_j}+Q_{H_d}\right)&=& - Q_S
 \left(\begin{array}{ccc}
  1 & 1& 1 \\
  1 & 1& 1 \\
  1 & 1& 1 \\
  \end{array} \right)_{ij}\;,\nonumber\\
 \left( Q_{L_i}+Q_{E^c_j}+Q_{H_d}\right)&=&
 \left(\begin{array}{ccc}
  -Q_S & -2 Q_{E^c_2}-4 Q_{L_2}- Q_S & -Q_{E^c_2}- 2 Q_{L_2} \\
  2 Q_{L_2} + 4 Q_{L_2} - Q_S & -Q_S & Q_{E^c_2} + 2 Q_{L_2} \\
  Q_{E^c_2} + 2 Q_{L_2} - Q_S & - Q_{E^c_2} - 2 Q_{L_2} - Q_S&  0 \\
  \end{array} \right)_{ij}\;.
\end{eqnarray}
It is clear that all of the up quarks get their masses from tree
level Yukawa interactions. On the other hand, none of the down-type
quarks are allowed to have tree level Yukawas, and only the tau
lepton is permitted to have a direct tree level mass. The massless
fermions are to obtain their masses from non-holomorphic terms via
(\ref{fmass}). To see if this really happens it is necessary to
examine the charge matrices determining the flavor structures of
the non-holomorphic couplings:
\begin{eqnarray}
\label{ctermtext}
 \left( Q_{Q_i}+Q_{U^c_j}-Q_{H_d}\right)&=&
Q_S  \left(\begin{array}{ccc}
  1 & 1& 1 \\
  1 & 1& 1 \\
  1 & 1& 1 \\
  \end{array} \right)_{ij}\;,\nonumber\\
 \left( Q_{Q_i}+Q_{D^c_j}-Q_{H_u}\right)&=&
 \left(\begin{array}{ccc}
  0 & 0& 0 \\
  0 & 0& 0 \\
  0 & 0& 0 \\
  \end{array} \right)_{ij}\;,\nonumber\\
 \left( Q_{L_i}+Q_{E^c_j}-Q_{H_u}\right)&=&
 \left(\begin{array}{ccc}
  0 & -2 Q_{E^c_2} - 4 Q_{L_2} & - Q_{E^c_2} - 2 Q_{L_2} + Q_S \\
  2 Q_{E^c_2} + 4 Q_{L_2} & 0  & Q_{E^c_2} + 2 Q_{L_2} + Q_S \\
  Q_{E^c_2} + 2 Q_{L_2} & - Q_{E^c_2} - 2 Q_{L_2}  & Q_S \\
  \end{array} \right)_{ij}\;.
\end{eqnarray}
Obviously, the up-type squarks are unable to develop any non-holomorphic
couplings: $C_U^{i j} = 0$ for all $i,j=1, 2, 3$. The situation
for down-type squarks is the opposite; they are allowed to develop
generic non-holomorphic trilinears with no texture zeroes: $C_D^{i
j} \neq 0$ for all $i,j$. The couplings of sleptons are
interesting; when $Q_{E^c_2}+2 Q_{L_2}\ne 0$ and $-Q_{E^c_2}-2Q_{L_2}+Q_S\ne 0$
they do not possess any flavor-changing non-holomorphic coupling:
$C_E^{i\neq j} = 0$ for all $i,j$. However, selectron and smuon
still have  non-holomorphic terms couplings. Consequently, the tau lepton acquires its
mass at tree level yet electron and muon obtain their masses via
(\ref{fmass}) with no leptonic FCNCs. We summarize the mechanisms of
mass generation for each fermion generation in Table
\ref{zp:tab:mass}.
\begin{table}[h]
\begin{center}
\begin{tabular}{|c|c|c|c|} \hline
      & 1st family & 2nd family  & 3rd family   \\ \hline
up-type quarks   & Y, $H_u$ & Y, $H_u$ & Y, $H_u$  \\ \hline
down-type quarks & R, $H_u$ & R, $H_u$ & R, $H_u$  \\ \hline
leptons          & R, $H_u$ & R, $H_u$ & Y, $H_d$  \\ \hline
\end{tabular}
\end{center}
\caption{The mechanisms for fermion mass generation: ``Y'' means
that mass is generated by tree level Yukawa interactions, and
``R'' means that the mass is generated radiatively via
(\ref{fmass}).} For each fermion, we also show that  which higgs
provides the vev for the corresponding fermion mass. \label{zp:tab:mass}
\end{table}

Given the allowed textures of Yukawa and non-holomorphic terms matrices
in (\ref{yukawatext}) and (\ref{ctermtext}), the effective
Yukawa interactions below the soft-breaking scale take the form
\begin{eqnarray}
\label{effYuk} -L_{eff} = h_u^{i j}\, (u_L)^c_i q_{j} H_u +
\tilde{h}_d^{i j}\, (d_L)^c_i q_j H_u^c  + \tilde{h}_e (e_L)^c L_1
H_u^c + \tilde{h}_{\mu} (\mu_L)^c L_2 H_u^c + {h}_{\tau}
(\tau_L)^{c} L_3 H_d
\end{eqnarray}
where the superscript $c$ stands for charge conjugation. The tilded
Yukawa couplings are generated by
non-holomorphic terms as in (\ref{fmass}): $\tilde{h}_d^{i j}
\propto C_D^{i j}$, $\tilde{h}_e \propto C_E^{1 1}$,
$\tilde{h}_{\mu} \propto C_E^{2 2}$. One notes that the tau lepton
is the only fermion which couples to $H_d$, in particular, it is vary
interesting that  the
entire quark sector behaves as in the SM (where $H_u$ serves as
the SM Higgs doublet $H_{SM}$) in contrast to its two-doublet
origin encoded in the superpotential (\ref{zp:eq:supo}). It is
clear from (\ref{effYuk}) that the entire hadronic FCNC is ruled
by the CKM matrix as in the SM, and no leptonic FCNC exists. In
this sense the family-nonuniversal \u1p model under consideration is
highly conservative  not only because of the minimality of the
spectrum but also because of the SM--like couplings of all
fermions but the tau lepton.

Since the model is already anomaly-free with minimal matter
content, SU(3)$_c$, SU(2)$_L$ and U(1)$_Y$ gauge couplings all
unify into a common value $g_0\simeq 1/\sqrt{2}$ at a scale
$M_{GUT}\approx 2 \times 10^{16}\ {\rm GeV}$ as in the MSSM. The
\u1p gauge coupling reads at the weak scale as
\begin{eqnarray}
g_1^{\prime\, 2}(M_Z) = \frac{g_0^2}{1 - 2 g_0^2 t_Z \mbox{Tr}[Q^2]}
\end{eqnarray}
where $t_Z= (4 \pi)^{-2} \log \left({M_Z}/{M_{GUT}}\right)$, and
clearly, $g_1^{\prime}(M_Z)$ depends on what values are assigned to the
independent charges $Q_{L_2}$, $Q_{E^c_2}$ and $Q_S$.

In the next section we will discuss some phenomenological implications of the minimal \u1p
model under consideration.

\section{Phenomenological Tests}\label{zp:sec:num}
In general, one can analyze the phenomenological implications of
our \u1p model as a function of the admissible
values ($e.g.$ $Q_S\neq 0$) of the charges $Q_{L_2}, Q_{E^c_2},
Q_S$. However, for simplicity we prefer to work with a
representative point in the space of \u1p charges and all other
model parameters. Therefore, we assign the following numerical
values to the free charges
\begin{eqnarray}
\label{numcharge} Q_{L_2}=2\;, \;\; Q_{E^c_2}=-3\;,\;\; Q_S=3
\end{eqnarray}
for which $g_1^{\prime}(M_Z)=0.196$ to be compared with
$g_Y(M_Z)=0.358$. With (\ref{numcharge}) the \u1p charges of
chiral fields get fixed to values depicted in Table
\ref{zp:tab:cha}. Note that the left-handed quarks are all
singlets under \u1p and right-handed up and down quarks are
charged oppositely under \u1p. Furthermore, the left-handed
electron does not couple to \zp.
\begin{table}[h]
\begin{center}
\begin{tabular}{|c|c|c|c|} \hline
             &  1st family & 2nd family & 3rd family \\ \hline
$Q_{Q_i}$    &  $0$    &  $0$   &  $0$  \\ \hline $Q_{U^c_i}$  &
$1$    &  $1$   &  $1$  \\ \hline $Q_{D^c_i}$  &  $-1$   &  $-1$ &
$-1$ \\ \hline $Q_{L_i}$    &  $0$    &  $2$   &  $1$  \\ \hline
$Q_{E^c_i}$  &  $-1$   &  $-3$  &  $1$  \\ \hline\hline
      & $Q_{H_u}$ & $Q_{H_d}$ & $Q_S$   \\ \hline
& $-1$ & $-2$ & $3$ \\ \hline
\end{tabular}
\end{center}
\caption{The \u1p charges of chiral fields corresponding to
the charge assignment in (\ref{numcharge}). \label{zp:tab:cha}}
\end{table}

\begin{table}[h]
\begin{center}
\begin{tabular}{|c|c|c|c|} \hline
             &  1st family & 2nd family & 3rd family \\ \hline
$Q_{Q_i}$    &  $1/3$    &  $1/3$   &  $1/3$  \\ \hline
$Q_{U^c_i}$ & $-1/3$    &  $-1/3$   &  $-1/3$  \\ \hline
$Q_{D^c_i}$ & $-1/3$ &  $-1/3$ & $-1/3$ \\ \hline $Q_{L_i}$    &
$0$ & $0$   & $0$  \\ \hline $Q_{E^c_i}$  &  $0$   &  $0$  &  $3$  \\
\hline\hline
      & $Q_{H_u}$ & $Q_{H_d}$ & $Q_S$   \\ \hline
& $0$ & $-3$ & $3$ \\ \hline
\end{tabular}
\end{center}
\caption{ An alternative charge assignment leading to an extremely
leptophobic \zp. \label{zp:tab:chax}}
\end{table}
Of course, there is no known fundamental reason for the
particular charge assignment in (\ref{numcharge}); one can adopt
some other numerical representation as well. Hence, as a distinct
case study consider another set of charges shown in Table \ref{zp:tab:chax}.
They satisfy all of the master relations in (\ref{noanomal}). In
fact, Table \ref{zp:tab:chax} has interesting properties in that
the \zp boson couples to no lepton but the right-handed tau lepton
and $\widehat{H}_u$ is neutral under \u1p. However, achieving such
an extremely leptophobic \zp boson has a price: the leptonic
Yukawa matrix and associated non-holomorphic terms are now allowed to have
nonvanishing off-diagonal entries, and thus the \zp boson
necessarily develops flavor-changing couplings to leptons which in
turn facilitate the leptonic FCNC decays $\mu \rightarrow e
\gamma$ or $\tau\rightarrow (\mu, e) \gamma$.
However the rates of these processes depend on the rotation matrix which
diagonalize the effective lepton Yukawa matrix. In the text we will not
pursue this option any further except to comment on it occasionally. We
will focus on the charge
assignments in Table \ref{zp:tab:cha} in discussing
phenomenological implications of the \zp boson.

In assigning numerical values to the rigid and soft parameters of
the theory we prefer to work at the weak scale. In fact, the
renormalization group flow is not needed at all as one can always
generate a given low-energy pattern from GUT scale parameters in
the absence of constraints like universality of the scalar soft
masses. Hence, we first fix the dominant Yukawa elements in the
superpotential to \be h_s=0.6\;,\;\; h_t=1.1 \ee for which their
RGEs develop no Landau pole up to $M_{GUT}$. Concerning the
soft-breaking sector, we choose gaugino masses and trilinear
couplings as in Table \ref{zp:tab:d1}, and scalar soft
mass--squares as in Table \ref{zp:tab:d2}. 
\begin{table}[h]
\begin{center}
\begin{tabular}{|c|c|c|c|c|c|c|c|c|} \hline
$M_1'$ & $M_1$ & $M_2$ & $M_3$&
  $A_S$ & $A_t$& $A_{\tau}$ & $C_b$ & $C_{\mu}$ \\ \hline
200 & 800 & 300 & 500 &
850   & 250  & 250        & 2000   & 1800       \\ \hline
\end{tabular}
\end{center}
\caption{ The gaugino masses and trilinear couplings at the
weak scale (in ${\rm GeV})$.} \label{zp:tab:d1}
\end{table}
\begin{table}[h]
\begin{center}
\begin{tabular}{|c|c|c|c|c|} \hline
$m_{H_u}^2$& $m_{H_d}^2$& $m_{S}^2$& $m_{Q,U^c,D^c}^2$& $m_{L,E^c}^2$ \\ \hline
 $-(175)^2$& $(823)^2$   & $-(565)^2$  & $(1000)^2$     & $(1400)^2$ \\ \hline
\end{tabular}
\end{center}
\caption{The soft mass--squared parameters (in ${\rm GeV}^2$) at
the weak scale.} \label{zp:tab:d2}
\end{table}
\newline
Notice that the negative
$m_S^2$ triggers the \u1p symmetry breaking. It is reasonable to expect
that by adjusting other soft SUSY breaking parameters one can get a positive
$m_S^2$ at the unification scale so that the \u1p symmetry is
radiatively broken, just like the radiative EWSB in the MSSM.
Investigating this possibility in detail is left for future work.
Also notice that in Table \ref{zp:tab:d1}, only the largest two 
non-holomorphic
terms, i.e. $C_b$ and $C_\mu$, are shown. As we already pointed out,
due to the fact that $C_f\propto m_f$ for the fermions whose masses are
due to the non-holomorphic terms, 
there is a hierarchy among the nonvanishing non-holomorphic terms,
 i.e. $m_b:m_s:m_d\approx C_b:C_s:C_d$ and $m_\mu:m_e\approx C_\mu:C_e$.
Since $C_\mu\gg C_e$, the left-right mixing in the smuon sector is much large
than the selectron sector, which tends to make $\tilde\mu_1$ lighter than
$\tilde e_1$. This may have interesting consequences for collider
signatures. For example, the chargino would more likely decay to
$\mu \nu_\mu\widetilde N_1$ than to $e\nu_e\widetilde N_1$. 

For the parameter values tabulated in Tables \ref{zp:tab:d1} and
\ref{zp:tab:d2}, the Higgs, $Z$, \zp and some of the fermion
masses turn out to be as in Table \ref{zp:tab:rev} for $\tan\beta=2$. 
\begin{table}[h]
\begin{center}
\begin{tabular}{|c|c|c|c|c|c|c|c|}\hline
$m_Z$& $m_{Z'}$& $m_{t}$& $m_b$ & $m_\mu$& $\alpha_{Z Z'}$&
  $m_h^{tree}$ \\ \hline
$91.2$ & $800$ & 175 & 2.9 & 0.101 &
$-2.76\times 10^{-3}$& $114.7$ \\ \hline
\end{tabular}
\end{center}
\caption{ Some particle masses (in ${\rm GeV}$) at the weak scale and
$Z-Z^{\prime}$ mixing angle for $\tan\beta =2$.} \label{zp:tab:rev}
\end{table}
\newline
Notice that the Higgs mass agrees with the LEP bounds
already at tree level.
Moreover, the \zp boson weighs nearly a ${\rm
TeV}$ and its mixing with the $Z$ boson, $\alpha_{Z Z'}$, remains
well inside the present experimental bounds. Furthermore, both $b$
quark and muon masses agree with experiments though they originate
from non-holomorphic terms rather than their Yukawa interactions with $H_d$.

Finally, for future use we also estimate the masses of the three
light neutralinos together with those of the stops, sbottoms and
smuons. The contributions from the D-terms associated with the \u1p are
taken into account in our calculation. The masses are shown in Table
\ref{zp:tab:rod}.   
\begin{table}[h]
\begin{center}\label{mutabzzm}
\begin{tabular}{|c|c|c|c|c|c|c|c|c|}\hline
$m_{\wtd \chi^0_1}$ &$m_{\wtd \chi^0_2}$& $m_{\wtd \chi^0_3}$&
$m_{\wtd t_1}$& $m_{\wtd t_2}$ &
$m_{\wtd b_1}$& $m_{\wtd b_2}$ &
$m_{\wtd \mu_1}$& $m_{\wtd \mu_2}$  \\ \hline
281 & 577 & 588 & 999 & 1051 & 783 & 1177 & 1318& 1725 \\ \hline
\end{tabular}
\end{center}
\caption{ The masses of light neutralinos and sfermions (in ${\rm GeV}$).
}\label{zp:tab:rod}
\end{table}
\newline
It is clear that LSP weighs $281 {\rm GeV}$ and light sbottom is the lightest
sfermion in the spectrum. Below the scale of \u1p breakdown the
model at hand resembles to the MSSM in that there is an effective
$\mu$ parameter induced: $\mu^{eff}=h_s\langle S\rangle=577\, {\rm
GeV}$ which lies right at the weak scale.

The numerical predictions above show that the \u1p model under
consideration does not have any obvious contradiction with the existing
phenomenological bounds. As part of the 'new physics search'
programme in laboratory and astrophysical environments,
establishing or excluding the class of models we are developing will require
analysis of various observables ranging from
Higgs boson signatures to dark matter in the universe. In the
following we will briefly discuss these observables, referring
to the numerical predictions above where needed.

\subsection{The Higgs Sector}\label{sec:higgssector}
In course of electroweak breaking $Z$ and \zp bosons acquire their
masses by eating, respectively, $\mbox{Im}\left[-\sin\beta H_u^0 +
\cos\beta H_d^0\right]$ and $\mbox{Im}\left[\cos\alpha \cos\beta
H_u^0 + \cos\alpha \sin\beta H_d^0-\sin\alpha S\right]$ where
$\cot\alpha = (v/\sqrt{2})$ $\sin\beta \cos\beta/\langle S
\rangle$ with $v^2/2 = \langle H_u^0 \rangle^2 + \langle H_d^0
\rangle^2$. The remaining neutral degrees of freedom  ${\cal{B}}=
\Big\{$ $\mbox{Re}\left[H_u^0\right]-\langle H_u^0 \rangle$,
$\mbox{Re}\left[H_d^0\right]-\langle H_d^0 \rangle$,
$\mbox{Re}\left[S\right]-\langle S \rangle$,
$\mbox{Im}\left[\sin\alpha \cos\beta H_u^0 + \sin\alpha \sin\beta
H_d^0+\cos\alpha S \right]$ $\Big\}$ span the space of massive
scalars. The physical Higgs bosons are given by $H_i =
{\cal{R}}_{i j} {\cal{B}}_{j}$ where the mixing matrix ${\cal{R}}$
necessarily satisfies ${\cal{R}} {\cal{R}}^T =1$, and it has
already been computed up to one loop order in
\cite{Cvetic:1997ky,everett,han,amini}. In the CP-conserving limit
the theory contains three CP-even, one CP-odd, and a charged Higgs
boson. The CP--odd scalar is typically heavy as its mass-squared
goes like $A_S \langle S \rangle $. It differs from the MSSM
spectrum by one extra CP-even scalar. At tree level, the lightest
Higgs mass is bounded as
\begin{eqnarray}
\label{mhupper} m_{H_1}^2 \leq M_Z^2 \cos^2 2 \beta + \frac{1}{2}
h_s^2 v^2 \sin^2 2\beta + g_1^{\prime\, 2} \left(Q_{H_d} \cos^2
\beta + Q_{H_u} \sin^2 \beta\right)^2 v^2
\end{eqnarray}
where the first term on the right-hand side is the MSSM
bound where the lightest Higgs is lighter than the $Z$ boson at
tree level. The second term is an $F$-term contribution that also
exists in the NMSSM \cite{NMSSM}. The last term, the \u1p D-term
contribution, enhances the upper bound in proportion to
$g_1^{\prime\, 2}$. Hence, rather generically the \u1p models
have a sufficiently large $m_{H_1}$ to higher
values making it likely that the lightest Higgs
lies beyond the LEP II kinematic reach.

Interestingly, the model favors specific values for $\tan\beta$
when $m_{Z'}$ approaches $m_Z$. Indeed, for such a light \zp
boson the $Z-Z^{\prime}$ mixing is suppressed by the
mixing mass-squared term $M_{Z-Z^{\prime}}^2 = (1/2)
g_1^{\prime} \sqrt{g_2^2 + g_Y^2}\, v^2 \left(Q_{H_d} \cos^2\beta
- Q_{H_u} \sin^2 \beta\right)$. This then requires $\tan\beta \sim
\sqrt{Q_{H_d}/Q_{H_u}}$, so $\tan\beta$ is completely determined by the
charge assignment! On the other hand, when \zp is sufficiently
heavy, this constraint on $\tan\beta$ is absent.

For the \u1p model example we analyze, the charge assignments in
Table \ref{zp:tab:cha} ensure that the \u1p D-term contribution to
the upper bound of the Higgs mass receives equal contributions
from $H_u$ and $H_d$. Moreover, for $\tan\beta \simeq \sqrt{2}$,
$Z-Z^{\prime}$ mixing would have been absent irrespective of the
scale of the \zp mass. Notably, if one switches to charge
assignments in Table \ref{zp:tab:chax} then the \u1p D-term
contribution to (\ref{mhupper}) gets significantly reduced at
large values of $\tan\beta$.

\subsection{The status of the fine-tuning problem}
One crucial message conveyed by the relative heaviness of the
lightest Higgs boson in \u1p models is that there is no need for
large radiative corrections in order to agree with the LEPII lower
bound. Indeed, when one-loop radiative corrections are included
the Higgs mass obeys the upper bound
\begin{eqnarray}
\label{mhupperx} \overline{m}_{H_1}^2 \leq m_{H_1}^{2} + \frac{3
m_t^4}{2 \pi^2 v^2}\, \log\frac{m_{\widetilde{t}}^2}{m_t^2}
\end{eqnarray}
where $m_{H_1}^{2}$ is the right hand side of Eq. (\ref{mhupper}).
The one-loop piece is an approximate result (note it does not depend
on $h_s$) that holds when ($i$) the loop contributions are
renormalized at $Q\sim m_{\widetilde{t}}$, ($ii$) all terms
involving the gauge couplings are neglected, and ($iii$) stop LR
mixing is much smaller than the diagonal terms such that the two
physical stops are nearly degenerate with mass $m_{\widetilde{t}}$
(see \cite{everett,amini} for exact results). The radiatively
corrected upper bound (\ref{mhupperx})  can be used to place a
lower bound on the stop mass
\begin{eqnarray}
m_{\widetilde{t}}\geq m_t\, e^{\left({\overline{m}_{H_1}^2}-
{m_{H_1}^{2}}\right)\, \frac{\pi^2 v^2}{3 m_t^4}}
\end{eqnarray}
where $v\approx 246\ {\rm GeV}$ is the electroweak breaking scale.
Consequently, when $\overline{m}_{H_1}=114\, {\rm GeV}$ the SUSY
breaking scale has the lower bounds $m_{\widetilde{t}}^2\simgt 3\
M_Z^2$ and $m_{\widetilde{t}}^2\simgt 4\ M_Z^2$ for parameter
values in Table \ref{zp:tab:cha} and Table \ref{zp:tab:chax},
respectively. A comparison of these results with the MSSM
expectation, $m_{\widetilde{t}}^2\simgt 50\ M_Z^2$ \cite{jose},
demonstrates that in \u1p models the SUSY breaking scale well be
close to the top mass. This result, which demonstrates the absence
of the little hierarchy problem in these class of models, stems
from the fact that the tree-level upper bound (\ref{mhupper}) is
already large enough to drag Higgs mass near the LEP lower bound.

The results above, however, should be taken with care. The main
reason is that the \zp boson should be heavy enough to satisfy
the bounds from precision data. In particular, the $Z$-$Z^{\prime
}$ mixing angle should be a few $\times 10^{-3}$, as mentioned and
computed before. This may occur because of a somewhat heavy \zp,
or because there exists a selection rule that enforces
approximately the interesting relation
$Q_{u}v_{u}^{2}-Q_{d}v_{d}^{2}\simeq 0$ (as in the 'large
trilinear vacuum' of \cite{Cvetic:1997ky,erler}).

We also want to mention that the recent analysis of the NMSSM
\cite{gunion} finds that fine-tuning \cite{fine} can be
significantly reduced especially in parameter regions with a light
pseudoscalar boson. The reason is that the invisible decay rate of
the Higgs boson gets enhanced (and thus it escapes detection at
LEP) via its decays into pairs of pseudoscalars.

\subsection{The \zp Couplings}
The \zp boson mixes with $Z_{\mu}=\cos\theta_W W^3_{\mu} - \sin\theta_{W} B_{\mu} $ after the electroweak
breaking since Higgs fields are charged under both U(1)$_Y$ and \u1p. On
top of this $B_{\mu}$ and \zp 
can exhibit kinetic mixing \cite{kolda}. In the presence of these
mixings the mass-eigenstate gauge bosons 
assume varying electroweak and \u1p components and these reflect
themselves in their interactions with matter 
species. For instance, the neutral vector boson observed in LEP experiments corresponds to
\begin{eqnarray}
Z^{(1)}_{\mu} = \cos \alpha_{Z Z^{\prime}} Z_{\mu} + \sin \alpha_{Z
Z^{\prime}} Z^{\prime}_{\mu} 
\end{eqnarray}
in the absence of kinetic mixing. The couplings of $Z^{(1)}_{\mu}$
to fermions deviate from their MSSM configuration in proportion
to $\alpha_{Z Z^{\prime}}$  and as a function of
$M_{Z_1}/M_{Z_2}$. All such \u1p impurities can be conveniently
represented by S, T and U parameters in a way useful for \zp
searches in electroweak precision data \cite{precision}.

In the following we will discuss the couplings of the \zp boson
rather than those of $Z^{(1)}_{\mu}$ or the heavy one
$Z^{(2)}_{\mu}$ as this is the crucial part of the information
needed for \u1p phenomenology. Depending on the mixing scheme,
kinetic or otherwise, one can always go to the physical basis for
gauge bosons by appropriate rotations. The \u1p charges of the
chiral fields shown in Table \ref{zp:tab:qn} are sufficient for
specifying their interactions with the \zp boson. The physical bases
for fermions are achieved by diagonalizing their Yukawa matrices
via the unitary transformations ${ h}_d^{diag}= V_R^{d} { h}_d
V_L^{d \dagger}$, ${ h}_u^{diag}= V_R^{u} { h}_u V_L^{u \dagger}$
and  ${ h}_e^{diag}= V_R^{e} { h}_e V_L^{e \dagger}$. Then the
physical fermions couple to \zp as $ g_1^{\prime} J_{\mu}\,
Z^{\prime\, \mu} + \mbox{h.c.}$ where
\begin{eqnarray}
J_{\mu}&=&\left[ \overline{d_L}\, \gamma_{\mu} V_L^{d} \left(\begin{array}{ccc}
Q_{Q_1}&0&0\\
0&Q_{Q_2}&0\\
0&0&Q_{Q_3}\end{array}\right) V_L^{d \dagger}\, d_L -
\overline{d_R}\, \gamma_{\mu} V_R^{d} \left(\begin{array}{ccc}
 Q_{D^c_1} &0&0\\
0&Q_{D^c_2}&0\\
0&0&Q_{D^c_3}\end{array}\right) V_R^{d \dagger}\, d_R \right]\nonumber\\
&+&\left[ \overline{u_L}\, \gamma_{\mu} V_L^{u} \left(\begin{array}{ccc}
Q_{Q_1}&0&0\\
0&Q_{Q_2}&0\\
0&0&Q_{Q_3}\end{array}\right) V_L^{u \dagger}\, u_L -
\overline{u_R}\, \gamma_{\mu} V_R^{u} \left(\begin{array}{ccc}
Q_{U^c_1}&0&0\\
0&Q_{U^c_2}&0\\
0&0&Q_{U^c_3}\end{array}\right) V_R^{u \dagger}\, u_R \right]\nonumber\\
&+&\left[ \overline{e_L}\, \gamma_{\mu} V_L^{e} \left(\begin{array}{ccc}
Q_{L_1}&0&0\\
0&Q_{L_2}&0\\
0&0&Q_{L_3}\end{array}\right) V_L^{e \dagger}\, e_L -
\overline{e_R}\, \gamma_{\mu} V_R^{e} \left(\begin{array}{ccc}
Q_{E^c_1}&0&0\\
0&Q_{E^c_2}&0\\
0&0&Q_{E^c_3}\end{array}\right) V_R^{e \dagger}\, e_R \right]
\end{eqnarray}
so that generically the \zp boson develops flavor-changing couplings if
there are intergenerational
mixings in the Yukawa matrices and/or if the \u1p charges are
family-dependent. A short glance at
the effective Yukawa interactions in (\ref{effYuk}) reveals that the
charged leptons are already
in their physical bases whereas the quarks exhibit nontrivial mixings
diagonalizations of which
induce flavor violation in charged-current vertices via $V_{CKM} =
V_{L}^{u} V_{L}^{d\, \dagger}$.
However, there are no flavor-changing \zp couplings to quarks at all.
The reason is that  \u1p charges of quarks are all family-universal
according to the anomaly-free
solutions in (\ref{noanomal}). In conclusion, the \zp boson couples to
fermions rather generically via
\begin{eqnarray}
\label{zpcurrent} J_{\mu} = \frac{1}{2} \sum_{i} \overline{\psi_i}
\gamma_{\mu} \left[ \left(Q_{left}^i - Q_{right}^i\right) -
\left(Q_{left}^i + Q_{right}^i\right) \gamma_5 \right] \psi_i
\end{eqnarray}
with no potential for tree-level flavor violation.

It is useful to discuss (\ref{zpcurrent}) in light of the charge
assignments in Table \ref{zp:tab:cha}. First of all, one
automatically concludes that $J_{\mu}$ is a $V+A$ current for
quarks, that is, each quark couples to $Z^{\prime}_{\mu}$ via
$\overline{q_R} \gamma^{\mu} q_{R}$ current only. In particular,
there is no involvement of the left-handed quark fields. On the
other hand, leptons possess varying vector and axial couplings due
to their family-nonuniversal \u1p charges. In fact,
$Z^{\prime}_{\mu}$ couples to the leptonic currents $(1/2)
\overline{e} \gamma_{\mu}( 1 + \gamma_5) e$, $(1/2) \overline{\mu}
\gamma_{\mu} (5 + \gamma_5) \mu$ and $- (1/2) \overline{\tau}
\gamma_{\mu} \gamma_{5} \tau$. Therefore, the electronic current is
purely right-handed as for quarks, the muonic current possesses a
sizeable vector part, and the tauonic current is purely
axial-vector type. Moreover, \zp boson does not couple to electron
neutrinos at all, and its coupling to the muon neutrino current is
twice larger than that to the tau neutrino current.  These
chirality and flavor sensitivities of the \u1p currents can have
important implications for \zp searches at colliders. If
one switches to charges assignments in Table \ref{zp:tab:chax} the
hadronic currents maintain their structure except for a resizing
by $1/3$, and the only surviving leptonic current turns out to be
that of the right-handed tau lepton. Consequently, this particular
charge assignment gives rise to an almost completely  leptophobic
\zp.

The kinetic terms of the Higgs fields completely determine the
couplings of \zp to Higgs bosons. In close similarity to $Z$ boson
couplings one can have vertices involving two \zp and two Higgs
bosons, or two \zp with a single CP-even Higgs boson, or a single
\zp accompanied by one CP-even and one CP-odd Higgs boson. A
single $Z^{\prime}_{\mu}$, for instance, couples to $H_i$ and
$H_j$ via $\left(p_{H_i}-p_{H_j}\right)^{\mu}$ times
\begin{eqnarray}
\label{coup1} 2 g_1^{\prime} \left({\cal{R}}\right)_{i 4} \left[
Q_{H_u} \cos\beta \sin\alpha \left({\cal{R}}\right)_{j 1} +
Q_{H_d} \cos\beta \sin\alpha \left({\cal{R}}\right)_{j 2} + Q_S
\cos\alpha \left({\cal{R}}\right)_{j 3}\right]
\end{eqnarray}
which vanishes unless $H_i$ and $H_j$ possess opposite CP
compositions. Unlike this, however, coupling of $H_i$ to
$Z^{\prime}_{\mu}\, Z^{\prime\, \mu}$ involves only its CP-even
component:
\begin{eqnarray}
\label{coup2} 2 g_{1}^{\prime\, 2} \left[ Q_{H_u}^2 \langle H_u^0
\rangle \left({\cal{R}}\right)_{i 1} + Q_{H_d}^2 \langle H_d^0
\rangle \left({\cal{R}}\right)_{i 2} + Q_{S}^2 \langle S \rangle
\left({\cal{R}}\right)_{i 3}\right]\; .
\end{eqnarray}
Finally, $H_i$ and $H_j$ couple to $Z^{\prime}_{\mu}\, Z^{\prime\,
\mu}$  via
\begin{eqnarray}
\label{coup3} &&g_1^{\prime\, 2} \Bigg[Q_{H_u}^2 \left\{
\left({\cal{R}}\right)_{i 1} \left({\cal{R}}\right)_{j 1} +
\cos^2\beta \sin^2\alpha  \left({\cal{R}}\right)_{i 4}
\left({\cal{R}}\right)_{j 4} \right\}\nonumber\\ &+& Q_{H_d}^2
\left\{ \left({\cal{R}}\right)_{i 2} \left({\cal{R}}\right)_{j 2}
+ \sin^2\beta \sin^2\alpha \left({\cal{R}}\right)_{i 4}
\left({\cal{R}}\right)_{j 4} \right\} \nonumber\\ &+&Q_{S}^2
\left\{ \left({\cal{R}}\right)_{i 3} \left({\cal{R}}\right)_{j 3}
+ \cos^2\alpha \left({\cal{R}}\right)_{i 4}
\left({\cal{R}}\right)_{j 4}\right\}\Bigg]\,.
\end{eqnarray}
The couplings of the Higgs bosons to distinct vector bosons,
$i.e.$ to $Z_{\mu}\, Z^{\prime}_{\nu}$, are obtained by picking up
both U(1)$_Y$ and \u1p contributions to Higgs kinetic terms.
Clearly, once $Z$ and \zp are rotated to their physical bases both
$Z_{\mu}\, Z_{\nu}$ and $Z^{\prime}_{\mu}\, Z^{\prime}_{\nu}$ type
structures will induce Higgs couplings to dissimilar vector bosons
via operators of the form $Z^{(1)}_{\mu}\, Z^{(2)}_{\nu}$.

The expressions for couplings presented above are general enough
to cover supersymmetric CP violation effects. In the CP-conserving
theory, as was assumed in constructing the soft-breaking sector in
Sec. 2, the Higgs bosons possess definite CP quantum numbers, in
particular, ${\cal{R}}_{4 i} =0$ for all $i\neq 4$ \cite{everett}.

\subsection{\zp Searches at Hadron Colliders}
>From a phenomenological point of view, the \u1p model under
concern differs from the MSSM by having one extra CP-even Higgs
boson, one extra neutral gauge boson, and two extra neutral
fermions. The ultimate confirmation of the model thus requires a
complete construction of all these states in laboratory or
astrophysical/cosmological environments. Here in this subsection
we will provide a rather brief description of \zp signatures
in accelerator experiments (See \cite{leike} for a review), in
particular, in hadron colliders $e.g.$ the Tevatron and upcoming
LHC. Needless to say, \zp signals at linear colliders are much
cleaner than at hadron machines but presently ILC is only being
planned (presumably as a post-LHC precision measurement
environment).

The LHC (Tevatron) is expected to probe \zp bosons as heavy as $4\
{\rm TeV}$ ($0.8\ {\rm TeV})$ depending on the model parameters,
on the luminosity reach of the collider, and on the size of
uncertainties coming from detector acceptances and systematic
errors \cite{leike,couplings1}.  \zp production proceeds via
various channels. It can be produced directly via quark--antiquark
fusion giving rise to $p\,p/\overline{p}\,p \rightarrow Z^{\prime}
X$ or indirectly via Higgs or $Z$ boson decays such as
$H_1\rightarrow Z^{\prime}\, Z^{\prime}$, $H_1 \rightarrow
Z^{\prime}\, H_4$ and $Z\rightarrow Z^{\prime}\, H_1$. Each of
these and similar contributions to \zp production can be analyzed
by using the expressions for the couplings given in
(\ref{zpcurrent}, \ref{coup1}, \ref{coup2}, \ref{coup3}) in Sec. 4.2
above. Among all these production channels the dominant one is the
quark-antiquark annihilation (at NLO in QCD gluon-quark scattering
into \zp is also important), and it facilitates direct $p\, p$ or
$\overline{p}\, p$ fusion into \zp. The produced \zp boson will
subsequently decay into leptons or jets. The latter are seldom
useful for \zp search due to large QCD background. The leptonic
signals, however, are particularly promising due to their good
momentum resolution and one's ability to suppress the MSSM
background at high dilepton invariant masses \cite{couplings1}.
When the subprocess center of mass energy $\simeq M_{Z^{\prime}}$
the \zp propagator resonates to give
\begin{eqnarray}
\label{sigmall} \sigma\left(p\,p \rightarrow
Z^{\prime}_{\searrow_{\ell^+ \ell^-}}\, X\right) =
\sigma\left(p\,p \rightarrow Z^{\prime}\, X\right)\,
\mbox{BR}\left(Z^{\prime} \rightarrow \ell^+\ell^-\right)
\end{eqnarray}
with a similar expression for $p\, \overline{p}$ collisions. Here
the \zp production rate is given by
\begin{eqnarray}
\sigma\left(p\,p \rightarrow Z^{\prime} X\right) &=&\sum_{q}
\frac{4 \pi^2}{3 s M_{Z^{\prime}}}\, \Gamma\left(Z^{\prime}
\rightarrow q\, \overline{q}\right)\;
\int_{\frac{M_{Z^{\prime}}^2}{s}}^{1} \frac{d x}{x}\nonumber\\
&\times& \left[ f_{\overline{q}}^{p}\left(x,
M_{Z^{\prime}}\right)\, f_{q}^{p}\left(\frac{M_{Z^{\prime}}^2}{x
s}, M_{Z^{\prime}}\right) + f_{{q}}^{p}\left(x,
M_{Z^{\prime}}\right)\,
f_{\overline{q}}^{p}\left(\frac{M_{Z^{\prime}}^2}{x s},
M_{Z^{\prime}}\right)\right]
\end{eqnarray}
where $f_x^y(a,b)$ stands for the probability of finding parton
$x$ in hadron $y$ with a momentum fraction $a$ at the relevant
energy scale $b$ of the scattering process. The partial fermionic
width of the \zp
\begin{eqnarray}
\label{gammall}
\Gamma\left(Z^{\prime} \rightarrow \psi_i
\overline{\psi}_i\right) = \frac{2 N_c}{3}\, {\alpha_1^{\prime}}\,
M_{Z^{\prime}}\, \left(Q_{left}^{i\, 2} + Q_{right}^{i\,
2}\right)\;,
\end{eqnarray}
as follows from (\ref{zpcurrent}), collects all model parameters
pertaining to the massless fermion sector. Presently, the CDF and
D0 experiments continue to explore \zp signatures by projecting
the measurement of (\ref{sigmall}) into possible values of
${\alpha_1^{\prime}} Q_{left}^{i\, 2} \mbox{BR}\left(Z^{\prime}
\rightarrow \ell^+\ell^-\right)$ in the plane of up and down quark
couplings \cite{cdfd0}.

For the minimal \u1p model under consideration, the following properties
could be important for collider searches for the \zp boson:
\begin{itemize}
\item At $e^+e^-$ (or future $\mu^+\mu^-$) colliders running above
the $Z$ pole the \zp effects can be parameterized in terms of
semi-electronic four-fermion operators. The scale of such
operators are ${\cal{O}}(10\ {\rm TeV})$ at LEP II. The combined
results of all four LEP collaborations \cite{lep} show that when
\zp couples to electrons of one chirality only (either to left or
right, not both) then bounds on $M_{Z^{\prime}}$ are rather weak.
This is indeed the case in our minimal \u1p model in which \zp
couples to the right-handed electron current only. Consequently,
it suffices to have $M_{Z^{\prime}} \simgt 0.7\ {\rm TeV}$ for LEP
II bounds to be respected. Clearly, if one switches to the charge
assignments in Table \ref{zp:tab:chax}, there is no LEP (or future
muon collider) bound to speak of (except for the precision
measurements at the $Z$ or \zp poles).

\item In the framework of the \u1p models under consideration, at hadron
colliders the \zp boson is produced by the fusion of right-handed
quarks. The decays of the produced \zp into leptons offer a rather
clean signal for experimental purposes \cite{couplings1,leike}. As
suggested by (\ref{gammall}) the larger the sum $Q_{left}^{i\, 2} +
Q_{right}^{i\, 2}$  larger the number of dilepton events.
Therefore, the number of $\mu^+\mu^-$ events must be 13 times
larger than $e^+e^-$ events and 26 times than $\tau^+\tau^-$
events. This rather strong preference for muon production gives a
clear signature of the model under concern. Of course, if one
switches to \u1p charges in Table \ref{zp:tab:chax} then \zp
effects show up only in the $\tau^+\tau^-$ production.

At hadron colliders, one of the most important observables is the
forward-backward asymmetry \cite{couplings1,leike}. It is a
measure of the angular distribution of the signal, and is
proportional to the vector and axial couplings of both the initial
and final state fermions in the process. For the \u1p charges in
Table \ref{zp:tab:cha} it vanishes for $\tau^+\tau^-$ production,
and is 5 times larger for $\mu^+\mu^-$ production than for
$e^+e^-$ signal. For the alternative charge assignments in Table
\ref{zp:tab:chax} there is no asymmetry at all; the signal is
distributed equally in forward and backward hemispheres.

In experiments with polarized proton beams one can define
spin-dependent asymmetries which probe chiral couplings of the
initial and final state fermions separately \cite{spin}. The
left-right asymmetry, defined with respect to the parent proton
helicity, is proportional to the multiplication of the vector and
axial couplings of the quarks, and it is universal for all quarks
in either of the charge assignments Tables \ref{zp:tab:cha} and
\ref{zp:tab:chax}. On the other hand, forward-backward asymmetry
for polarized protons measures the chiral couplings of the leptons
in isolation in a way similar to the forward-backward asymmetry of
unpolarized beams.
\end{itemize}

In this subsection we have discussed very briefly the prospects
for \zp searches at colliders within our minimal \u1p extension of
the MSSM. Clearly, for a complete determination of the \zp
signatures it is necessary to perform a detailed study of all
relevant processes. Notice that the particular model we showed at the 
beginning of this section has a 
\zp at 800 GeV. One can certainly lower the \zp mass to increase the
chance of detectability at the Tevatron. Smaller \zp mass will typically
increase the $Z$-\zp mixing angle. But as shown in section
\ref{sec:higgssector}, special values of $\tan\beta$ can be chosen to
reduce the mixing. We have found that it is possible to make \zp as
light as around $500$ GeV and the mixing angle close to the border line of
the experimental bound.

\subsection{The Neutrino Masses}
By construction, the model analyzed in this work does not contain
any fields necessary for inducing the neutrino masses and mixings.
These can be generated via various mechanisms \cite{smirnov,langx,
ma,kanex}. For a consistent analysis of the neutrino sector one has to import
appropriate fields into the spectrum and analyze their consequences,
especially for anomaly cancellation. Here, we simply take the see-saw contribution
\be
\Delta W_{\nu}=Y_\nu^{ij}\frac{L^i H_u L^j H_u}{M}
\ee
to the superpotential as a basis for our brief discussion.
Here $Y_\nu^{ij}$ are some $\mathcal{O}(1)$ couplings, and
$M$ represents the Majorana mass scale.  In models with
additional \u1p symmetry, some of the entries
of $Y_{ij}$ could be forbidden by the \u1p symmetry.
Indeed, a short glance at the charge assignments in
Table \ref{zp:tab:cha} reveals that $Y^{ij}_\nu$
should take the following form
\be
Y_\nu^{ij}=\left(
\begin{array}{ccc}
0 & a & 0 \\
a & 0 & 0 \\
0 & 0 & b
\end{array} \right)
\ee
where $a$ and $b$ are some coefficients. Clearly, this texture
does not account for the observed oscillation data,
and one has to invent some other way of inducing a viable $Y_\nu$.

On the other hand, for the \u1p charge assignments in Table
\ref{zp:tab:chax} the see-saw mechanism alone suffices to induce
all neutrino masses and mixings in full generality (at the expense
of opening up the lepton flavor violation effects). Analyzing
these patterns and constraints is left for further work.

\subsection{The Muon $g-2$}
We have already provided general expressions for $g_{\mu}-2$
in Sec. 2. Thanks to the non-holomorphic operator
$C^{22}_E H_u^* \wtd L^2 \wtd E_R^{c2}$ one can induce both
muon mass and $g_{\mu}-2$ via one-loop neutralino-smuon
diagram. On the other hand, there is no similar chirality-flip
operator on the $\wtd\nu_\mu$ line so that the chargino
contribution is a two-loop effect and is thus negligible. Inducing
the muon mass without violating $g_{\mu}-2$ bounds is an
important constraint \cite{Borzumati:1999sp}, and for the parameter values listed before
we find
\be
a_\mu^{SUSY}=22 \times 10^{-10}
\ee
by using (\ref{zp:eq:gmu}). This result is well inside
the allowed room for 'new physics' contribution to muon
anomalous magnetic moment \cite{gm2exp}.

\subsection{The Cold Dark Matter}
The mapping of the CMB anisotropy provides precise information
about the densities of matter and dark energy in the universe. It
is now known with good precision that the matter distribution is
dominated by a non-baryonic non-relativistic component whose
candidate particle should be massive, stable, neutral and weakly
interacting.  Supersymmetric models with conserved $R$ parity
provide a natural candidate for cold dark matter (CDM) in the
lightest superpartner $i.e.$ the lightest neutralino $\chi_1^0$.
For the parameter values listed in Table \ref{zp:tab:d1} the LSP
turns out to be wino dominated \be \wtd\chi^0_1=-0.015 \wtd
Z'-0.019 \wtd B +0.967 \wtd W-0.197\wtd H_d+0.158\wtd H_d-0.004
\wtd S \ee with a rather small singlino component. For wino LSPs,
coannihilation during the freeze out is highly efficient. In fact,
the neutralino relic density turns out to be $\Omega_{\chi} h^2
\simeq 0.5\times 10^{-2}$ which is smaller than the observed CDM
density by an order of magnitude. Hence, as pointed out before
\cite{wino}, the wino LSP is far from being a viable CDM
candidate. However, non-thermal production can provide the actual
relic density, e.g., for the wino LSP, and decays of the moduli
fields into gauginos can help in enhancing $\Omega_{\chi}$ for
saturating the correct value of $\Omega_{CDM}$ \cite{randall}.
Clearly, if the LSP is dominated by other components $i.e.$
singlino, bino or $Z$'ino then one can saturate the observed value
of $\Omega_{CDM}$ since their annihilation rates are relatively
smaller than those of the Winos \cite{barger}.

\section{Conclusion}\label{zp:sec:con}
We have discussed ways of constructing an anomaly-free \u1p model (as
needed for solving the $\mu$ problem and moderating the fine-tuning
problem) with minimal matter content in order to maintain the
unification of gauge couplings . We have found and illustrated with some 
numerical examples that it is possible to achieve the cancellation of anomalies
with no exotic matter by invoking ($i$) family-nonuniversal \u1p charge
assignments and ($ii$) non-holomorphic soft-breaking operators.

The model discussed in this work is an anomaly-free version
of the generic \u1p model analyzed in \cite{Cvetic:1997ky}.
Indeed, the two models have identical matter spectrum. However,
achieving anomaly freedom without exotic states requires the
introduction of family-dependent \u1p charge assignments plus
non-holomorphic soft-breaking terms. Of course, \u1p models that
follow from E$_6$ breaking are anomaly-free thanks to the exotic
states present in the light spectrum \cite{wang}. In this sense,
the model discussed here constitutes an anomaly-free minimal \u1p
model.

>From the experimental point of view, distinguishing the minimal
\u1p model here from other \u1p models or from the MSSM requires
measurement of a number of observables. In general, establishing
the existence of a \u1p-extended MSSM structure necessitates
experimental evidence for \zp boson, extra Higgs bosons and extra
neutralino states. On the other hand, one might interpret certain
phenomenological results as being evidences for an extended gauge
sector. For instance, the EDM constraints generically require the
phase of the $\mu$ parameter (in the MSSM) to be rather small, and
this result can be naturally tied to the radiative nature of
the $\mu$ parameter in \u1p models \cite{everett}.

Distinguishing the minimal \u1p model here from other \u1p models
in the literature requires certain signatures which could come from
non-holomorphicity and family-dependent nature of the \zp
couplings. Concerning the latter, one recalls from Sec. 4.4 that
\zp decays into a specific difermion state, $e.g.$ $\mu^+\mu^-$,
can be significantly enhanced compared to others due to the family
dependence of the \u1p couplings displayed in Table \ref{zp:tab:cha}.
In fact, the quarks which participate in production and hadronic decays of 
the \zp boson are right-handed more often than is typical. These are
signals that  
cannot be found in other \u1p models. The family non-universality implies 
several collider events that enable one to  distinguish the minimal \u1p 
here from other models.

Being another important effect of family nonuniversality, one notes from  
Table  \ref{zp:tab:qn} that \zp does not couple to left-handed squarks 
and left-handed selectron, at all. In fact, its strongest coupling is to 
smouns, in particular, to the right-handed smuon. The  dominance 
of the right-handed currents (except the stau states) is interesting since 
right-handed sfermions (of the first two generations, especially) decay 
preferably into bino and right-handed fermions. In particular, multilepton
plus jet plus missing energy signals coming from left-handed squarks are 
now reduced. Besides these, dominance of the muon signal 
compared to others is a signal of the violation of lepton universality, 
and the \zp boson of Table  \ref{zp:tab:cha} could be a viable source of 
this.

The non-holomorphicity of the soft-breaking terms affect certain
observables in a distinct way. For instance, due to their
radiative origin the Higgs-fermion couplings depend on the
momentum transfer in a given scattering process, and thus,
non-holomorphic structures may be tested by measuring various
Higgs  branching fractions into fermions \cite{Borzumati:1999sp}.
Furthermore, the electric dipole moments (though not analyzed
here) are naturally suppressed since dipole moments are aligned
towards the fermion masses \cite{Borzumati:1999sp}. Finally,
the heavier the fermion larger the non-holomorphic trilinear, and
hence, the sfermion left-right mixings are enhanced for relatively heavy
fermions whose masses are due the the non-holomorphic terms.

In conclusion, we have analyzed the conditions for and
phenomenological consequences of  canceling the anomalies in \u1p
models with minimal matter content. We have briefly discussed a
number of observables ranging from fermion masses to dark matter
in the universe. The model explored here is minimal in that it is
a direct \u1p gauging of the MSSM plus a gauge singlet, and it
needs to be extended to include right-handed neutrinos to induce
neutrino masses and mixings. Moreover, the numerical examples
provided here can be extended to a sufficiently dense sampling of
the parameter space for determining the laboratory and
astrophysical implications of the model.

\section{Acknowledgements}
The authors thank Lisa Everett, Brent Nelson and James Wells for useful
discussions. We appreciate interesting correspondence with Pierre Fayet.
The work of 
G.K. and T.W. is supported by US Department of Energy. D.D. would
like to thank Michigan Center for Theoretical Physics for generous
hospitality while this work was started, Turkish Academy of
Sciences for financial support through GEBIP grant, and 
Scientific and Technical Research Council of Turkey for financial
support through project 104T503.


\begin{thebibliography}{99}
\bibitem{muprob}
P.~Fayet,
Phys.\ Lett.\ B {\bf 69}, 489 (1977).
J.~E.~Kim and H.~P.~Nilles,
Phys.\ Lett.\ B {\bf 138}, 150 (1984);
D.~Suematsu and Y.~Yamagishi,
Int.\ J.\ Mod.\ Phys.\ A {\bf 10}, 4521 (1995)
[arXiv:hep-ph/9411239];
M.~Cvetic and P.~Langacker,
Mod.\ Phys.\ Lett.\ A {\bf 11}, 1247 (1996)
[arXiv:hep-ph/9602424];
V.~Jain and R.~Shrock,
arXiv:hep-ph/9507238;
Y.~Nir,
Phys.\ Lett.\ B {\bf 354}, 107 (1995)
[arXiv:hep-ph/9504312].

\bibitem{gut}
J.~L.~Hewett and T.~G.~Rizzo,
  Phys.\ Rept.\  {\bf 183}, 193 (1989).

\bibitem{string}
M.~Cvetic and P.~Langacker,
Phys.\ Rev.\ D {\bf 54}, 3570 (1996) [arXiv:hep-ph/9511378].

\bibitem{dynamic}
C.~T.~Hill and E.~H.~Simmons,
  Phys.\ Rept.\  {\bf 381}, 235 (2003)
  [Erratum-ibid.\  {\bf 390}, 553 (2004)]
  [arXiv:hep-ph/0203079].

\bibitem{Cvetic:1997ky}
M.~Cvetic, D.~A.~Demir, J.~R.~Espinosa, L.~L.~Everett and P.~Langacker,
Phys.\ Rev.\ D {\bf 56}, 2861 (1997)
[Erratum-ibid.\ D {\bf 58}, 119905 (1998)]
[arXiv:hep-ph/9703317].

\bibitem{everett}
D.~A.~Demir and L.~L.~Everett,
Phys.\ Rev.\ D {\bf 69}, 015008 (2004)
[arXiv:hep-ph/0306240].

\bibitem{han}
T.~Han, P.~Langacker and B.~McElrath,
arXiv:hep-ph/0402064;
arXiv:hep-ph/0405244.

\bibitem{tianjun}
J.~Kang, P.~Langacker, T.~j.~Li and T.~Liu,
arXiv:hep-ph/0402086.


\bibitem{Erler:2000wu}
H.~C.~Cheng, B.~A.~Dobrescu and K.~T.~Matchev,
Phys.\ Lett.\ B {\bf 439}, 301 (1998)
[arXiv:hep-ph/9807246].
H.~C.~Cheng, B.~A.~Dobrescu and K.~T.~Matchev,
Nucl.\ Phys.\ B {\bf 543}, 47 (1999)
[arXiv:hep-ph/9811316].
J.~Erler,
Nucl.\ Phys.\ B {\bf 586}, 73 (2000)
[arXiv:hep-ph/0006051].

\bibitem{Langacker:2000ju}
P.~Langacker and M.~Plumacher,
Phys.\ Rev.\ D {\bf 62}, 013006 (2000)
[arXiv:hep-ph/0001204];
V.~Barger, C.~W.~Chiang, P.~Langacker and H.~S.~Lee,
Phys.\ Lett.\ B {\bf 580}, 186 (2004)
[arXiv:hep-ph/0310073].

\bibitem{Jack:2003pb}
  I.~Jack, D.~R.~T.~Jones and R.~Wild,
  Phys.\ Lett.\ B {\bf 580}, 72 (2004)
  [arXiv:hep-ph/0309165].


\bibitem{Hall:1990ac}
L.~J.~Hall and L.~Randall,
Phys.\ Rev.\ Lett.\  {\bf 65}, 2939 (1990).


\bibitem{Borzumati:1999sp}
F.~Borzumati, G.~R.~Farrar, N.~Polonsky and S.~Thomas,
Nucl.\ Phys.\ B {\bf 555}, 53 (1999)
[arXiv:hep-ph/9902443].


\bibitem{Camara:2003ku}
P.~G.~Camara, L.~E.~Ibanez and A.~M.~Uranga,
Nucl.\ Phys.\ B {\bf 689}, 195 (2004) [arXiv:hep-th/0311241].

\bibitem{ma1}
E.~Ma and M.~Raidal,
  Phys.\ Lett.\ B {\bf 491}, 297 (2000)
  [arXiv:hep-ph/0006253];
M.~A.~Cakir, S.~Mutlu and L.~Solmaz,
  arXiv:hep-ph/0501286.


\bibitem{jack}
I.~Jack and D.~R.~T.~Jones,
  Phys.\ Lett.\ B {\bf 457}, 101 (1999)
  [arXiv:hep-ph/9903365];
Phys.\ Rev.\ D {\bf 61}, 095002 (2000)
  [arXiv:hep-ph/9909570].

\bibitem{nonholo}
D.~M.~Capper,
J.\ Phys.\ G {\bf 3}, 731 (1977);
L.~Girardello and M.~T.~Grisaru,
Nucl.\ Phys.\ B {\bf 194}, 65 (1982).

\bibitem{martinex}
S.~P.~Martin,
  Phys.\ Rev.\ D {\bf 61}, 035004 (2000)
  [arXiv:hep-ph/9907550].


\bibitem{nima}
N.~Arkani-Hamed and R.~Rattazzi,
  Phys.\ Lett.\ B {\bf 454}, 290 (1999)
  [arXiv:hep-th/9804068].


\bibitem{buch}
B.~de Wit, M.~T.~Grisaru and M.~Rocek,
  Phys.\ Lett.\ B {\bf 374}, 297 (1996)
  [arXiv:hep-th/9601115];
D.~Bellisai, F.~Fucito, M.~Matone and G.~Travaglini,
  Phys.\ Rev.\ D {\bf 56}, 5218 (1997)
  [arXiv:hep-th/9706099];
M.~Dine and N.~Seiberg,
  Phys.\ Lett.\ B {\bf 409}, 239 (1997)
  [arXiv:hep-th/9705057];
F.~Gonzalez-Rey and M.~Rocek,
  Phys.\ Lett.\ B {\bf 434}, 303 (1998)
  [arXiv:hep-th/9804010];
E.~I.~Buchbinder, I.~L.~Buchbinder and S.~M.~Kuzenko,
  Phys.\ Lett.\ B {\bf 446}, 216 (1999)
  [arXiv:hep-th/9810239].

\bibitem{murayama}
J.~L.~Diaz-Cruz, H.~Murayama and A.~Pierce,
  Phys.\ Rev.\ D {\bf 65}, 075011 (2002)
  [arXiv:hep-ph/0012275].


\bibitem{lfv}
F.~Deppisch, J.~Kalinowski, H.~Pas, A.~Redelbach and R.~Ruckl,
arXiv:hep-ph/0401243.



\bibitem{amini}
D.~A.~Demir and N.~K.~Pak,
Phys.\ Rev.\ D {\bf 57}, 6609 (1998)
[arXiv:hep-ph/9809357];
H.~Amini,
New J.\ Phys.\  {\bf 5} (2003) 49
[arXiv:hep-ph/0210086].

\bibitem{NMSSM}
See for instance:\\
J.~R.~Ellis, J.~F.~Gunion, H.~E.~Haber, L.~Roszkowski and
F.~Zwirner,
Phys.\ Rev.\ D {\bf 39}, 844 (1989).
Y.~Daikoku and D.~Suematsu,
Phys.\ Rev.\ D {\bf 62}, 095006 (2000)
[arXiv:hep-ph/0003205].
D.~J.~Miller, R.~Nevzorov and P.~M.~Zerwas,
Nucl.\ Phys.\ B {\bf 681}, 3 (2004)
[arXiv:hep-ph/0304049].



\bibitem{jose}
J.~A.~Casas, J.~R.~Espinosa and I.~Hidalgo,
JHEP {\bf 0401}, 008 (2004)
[arXiv:hep-ph/0310137].

\bibitem{erler}
J.~Erler, P.~Langacker and T.~j.~Li,
  Phys.\ Rev.\ D {\bf 66}, 015002 (2002)
  [arXiv:hep-ph/0205001].


\bibitem{fine}
R.~Barbieri and G.~F.~Giudice,
Nucl.\ Phys.\ B {\bf 306}, 63 (1988);
B.~de Carlos and J.~A.~Casas,
Phys.\ Lett.\ B {\bf 309}, 320 (1993)
[arXiv:hep-ph/9303291];
G.~W.~Anderson and D.~J.~Castano,
Phys.\ Lett.\ B {\bf 347}, 300 (1995)
[arXiv:hep-ph/9409419];
Phys.\ Rev.\ D {\bf 52} (1995) 1693
[arXiv:hep-ph/9412322].
P.~H.~Chankowski, J.~R.~Ellis and S.~Pokorski,
Phys.\ Lett.\ B {\bf 423}, 327 (1998)
[arXiv:hep-ph/9712234];
R.~Barbieri and A.~Strumia,
Phys.\ Lett.\ B {\bf 433}, 63 (1998)
[arXiv:hep-ph/9801353];
P.~H.~Chankowski, J.~R.~Ellis, M.~Olechowski and S.~Pokorski,
Nucl.\ Phys.\ B {\bf 544}, 39 (1999)
[arXiv:hep-ph/9808275];
G.~L.~Kane and S.~F.~King,
Phys.\ Lett.\ B {\bf 451}, 113 (1999)
[arXiv:hep-ph/9810374];
M.~Bastero-Gil, G.~L.~Kane and S.~F.~King,
Phys.\ Lett.\ B {\bf 474}, 103 (2000)
[arXiv:hep-ph/9910506].
G.~L.~Kane, J.~Lykken, B.~D.~Nelson and L.~T.~Wang,
Phys.\ Lett.\ B {\bf 551}, 146 (2003)
[arXiv:hep-ph/0207168];
D.~A.~Demir,
  arXiv:hep-ph/0408043;
A.~Maloney, A.~Pierce and J.~G.~Wacker,
arXiv:hep-ph/0409127.

\bibitem{gunion}
R.~Dermisek and J.~F.~Gunion,
  arXiv:hep-ph/0502105.


\bibitem{kolda}
K.~S.~Babu, C.~F.~Kolda and J.~March-Russell,
Phys.\ Rev.\ D {\bf 57}, 6788 (1998)
[arXiv:hep-ph/9710441].
D.~Suematsu,
Phys.\ Rev.\ D {\bf 59}, 055017 (1999)
[arXiv:hep-ph/9808409].



\bibitem{precision}
J.~Erler and P.~Langacker,
Phys.\ Rev.\ Lett.\  {\bf 84}, 212 (2000)
[arXiv:hep-ph/9910315];
arXiv:hep-ph/0407097.

\bibitem{leike}
A.~Leike,
Phys.\ Rept.\  {\bf 317}, 143 (1999) [arXiv:hep-ph/9805494].

\bibitem{couplings1}
F.~del Aguila, M.~Cvetic and P.~Langacker,
Phys.\ Rev.\ D {\bf 48}, 969 (1993) [arXiv:hep-ph/9303299];
Phys.\ Rev.\ D {\bf 52}, 37 (1995) [arXiv:hep-ph/9501390];
F.~Del Aguila and M.~Cvetic,
Phys.\ Rev.\ D {\bf 50}, 3158 (1994) [arXiv:hep-ph/9312329];
A.~Leike,
Phys.\ Lett.\ B {\bf 402}, 374 (1997) [arXiv:hep-ph/9703263];
T.~Appelquist, B.~A.~Dobrescu and A.~R.~Hopper,
Phys.\ Rev.\ D {\bf 68}, 035012 (2003) [arXiv:hep-ph/0212073];
M.~Carena, A.~Daleo, B.~A.~Dobrescu and T.~M.~P.~Tait,
arXiv:hep-ph/0408098.

\bibitem{cdfd0}
[D0 Collaboration],  4577-Conf.; [The CDF and D0 Collaborations],
FERMILAB-CONF-04/179-E.

\bibitem{lep}
[LEP Collaboration],
arXiv:hep-ex/0312023.


\bibitem{spin}
A.~Fiandrino and P.~Taxil,
Phys.\ Rev.\ D {\bf 44}, 3490 (1991);
P.~Taxil, E.~Tugcu and J.~M.~Virey,
Eur.\ Phys.\ J.\ C {\bf 24}, 149 (2002) [arXiv:hep-ph/0111242].

\bibitem{smirnov}
A.~Y.~Smirnov,
arXiv:hep-ph/0411194.


\bibitem{langx}
P.~Langacker,
Phys.\ Rev.\ D {\bf 58}, 093017 (1998) [arXiv:hep-ph/9805281];
J.~Kang, P.~Langacker and T.~Li,
arXiv:hep-ph/0411404.

\bibitem{ma}
T.~Hambye, E.~Ma and U.~Sarkar,
Nucl.\ Phys.\ B {\bf 590}, 429 (2000) [arXiv:hep-ph/0006173].

\bibitem{kanex}
J.~Giedt, G.~L.~Kane, P.~Langacker and B.~D.~Nelson,
arXiv:hep-th/0502032.

\bibitem{gm2exp}
B.~L.~Roberts  [Muon g-2 Collaboration],
arXiv:hep-ex/0501012.

\bibitem{wino}
G.~F.~Giudice, M.~A.~Luty, H.~Murayama and R.~Rattazzi,
JHEP {\bf 9812}, 027 (1998)
[arXiv:hep-ph/9810442].


\bibitem{randall}
T.~Moroi and L.~Randall,
Nucl.\ Phys.\ B {\bf 570}, 455 (2000)
[arXiv:hep-ph/9906527];

\bibitem{barger}
V.~Barger, C.~Kao, P.~Langacker and H.~S.~Lee,
Phys.\ Lett.\ B {\bf 600}, 104 (2004)
[arXiv:hep-ph/0408120].

\bibitem{wang}
P.~Langacker and J.~Wang,
  Phys.\ Rev.\ D {\bf 58}, 115010 (1998)
  [arXiv:hep-ph/9804428];
 J.~Kang and P.~Langacker,
  Phys.\ Rev.\ D {\bf 71}, 035014 (2005)
  [arXiv:hep-ph/0412190].
\end{thebibliography}
\end{document}